\documentclass[10pt,letterpaper]{article}
\usepackage[margin=2.5cm]{geometry}
\usepackage[affil-it]{authblk} 
\usepackage[utf8]{inputenc} 
\usepackage{float}
\usepackage{relsize}
\usepackage{amsmath}
\usepackage{amssymb,breqn,physics}  
\usepackage{graphicx}
\usepackage{xcolor}
\usepackage{caption}

\usepackage{fancyhdr}
\usepackage{abstract}

\numberwithin{equation}{section}

\usepackage[hidelinks]{hyperref}
\hypersetup{
    citecolor=black, 
    colorlinks = true,
    urlcolor={black},
    linkcolor=.  
    }

\makeatother
\definecolor{hyperlink}{rgb}{0,0,95}  
\usepackage[
    backend=biber,
    style=numeric,
    sorting=none 
]{biblatex}
\title{\bfseries{ Imprints of Anisotropy on the Power Spectrum in Matter Dominated Bouncing Universe as Background}}
\author{Asha B Modan\footnote{Electronic address: \href{mailto:ashabmodan98@gmail.com}{ ashabmodan98@gmail.com} }, Sukanta Panda\footnote{Electronic address: \href{mailto:sukanta@iiserb.ac.in}{ sukanta@iiserb.ac.in} }, Arun Rana\footnote{Electronic address: \href{mailto:arunrana@iiserb.ac.in}{ arunrana@iiserb.ac.in}}}
\date{\today}
\affil{Department of Physics, Indian Institute of Science Education and Research Bhopal, Bhopal, Madhya Pradesh, India}

\renewcommand{\thesection}{\Roman{section}.} 
\renewcommand{\thesubsection}{\Alph{subsection}.} 
\usepackage[explicit]{titlesec}
\titleformat{\section}
{\normalfont\normalsize\bfseries}{\thesection}{12pt}{\filcenter
\MakeUppercase{#1}} 
\titleformat{\subsection}
{\normalfont\normalsize\bfseries}{\thesubsection}{12pt}{\filcenter{#1}} 

\begin{document}
\captionsetup[figure]{labelfont={up},labelformat={default},labelsep=colon, name={FIG.}} 

\maketitle 
\begin{abstract}
 
 In this paper, we aim to investigate the effects of the anisotropy on the scale-invariant power spectrum considering the matter-dominated collapsing universe as background and look for the deviations from the scale invariance. Having set up this background, we consider a massless scalar field and work out the correlations, first by using the perturbative approach in which the anisotropic background is approximated with an effective isotropic metric represented by the metric of matter dominated collapsing universe, second by directly solving the field equation numerically, and then obtain the power spectrum for the range of modes which are of cosmological interest. Using both techniques, we get an upper bound on the deviation in the power spectrum from the scale invariance. We also work out the power spectrum for much smaller modes and look at whether it is possible to explain the observed anomalies in CMB via the matter bounce scenario. 

\end{abstract}
\vspace{1em} 

\section{Introduction}
The beginning of the universe has been a long-standing question for cosmologists. We have always wondered whether the universe began with a singularity or not. The Big Bang paradigm that consists of an initial singularity seems to be the most natural thought, as imagining a point-sized universe with an infinite density and temperature with all the fundamental interactions unified by a yet unknown framework is the most accessible course of action. Nonetheless, no one can exclude the possibility of a cyclical cosmological evolution of the universe where its size never shrinks to zero. Even quantum cosmologies support this latter perspective \cite{Nojiri:2017ncd}. The observations of cosmic microwave background (CMB) and the large scale structure \cite{WMAP:2003elm,Planck:2018vyg} strongly support the argument that the primordial curvature perturbations which generate nearly scale-invariant power spectrum act as seeds for the structures of our universe. This spectrum is the result of correlations between the quantum vacuum fluctuations \cite{Mukhanov:1981xt,Starobinsky:1979ty,Press:1980zz} during the cosmic inflation; an era of exponentially accelerated expansion of our universe right after the Big Bang \cite{Guth:1980zm,Linde:1981mu}. Not just the inflation, a corresponding scale-invariant power spectrum is obtained in the case of a matter-dominated collapsing universe, as pointed out in \cite{Wands_1999,Finelli:2001sr}. This kind of model, which is succeeded by a non-singular bounce, is called the matter bounce scenario and can act as an alternative to the inflation generating the observed power spectrum of the primordial fluctuations \cite{Brandenberger:2009jq,Brandenberger:2012zb}. To understand the transition between the expanding and contracting phases of the universe, new physics is required. This transition can be singular if the  ekpyrotic scenario \cite{Khoury:2001wf} is to be considered or non-singular. For the latter case various methods have been developed, for example, the modifications of gravitational action in torsion gravity \cite{Poplawski:2011jz,Cai:2011tc}, Horava-Lifshitz gravity \cite{Brandenberger:2009yt,Kiritsis:2009sh,Calcagni:2009ar}  or by introducing the Galileon field \cite{Qiu:2011cy,Easson:2011zy} or a ghost condensate \cite{Lin:2010pf,Buchbinder:2007ad,Creminelli:2007aq} consisting of matter that violates the positive energy condition. For a more detailed review, one can look at \cite{Novello:2008ra}.

Now, one of the main issues with bouncing cosmologies is the
anisotropic instabilities, known as 
Belinsky-Khalatnikov-Lifshitz (BKL) 
instability \cite{Belinsky:1970ew}. It occurs because, during the contracting phase, the rate at which the energy densities of the dust and radiation matter field increase is much lesser than the rate of increase in the energy densities contributed by the back-reaction of the anisotropies. Hence, it becomes necessary that
to have a bounce that is nearly isotropic, one has to fine-tune the initial conditions to such accuracies that the anisotropies never dominate. However, there is another way to resolve this problem, that is by introducing a scalar field accompanied by a steep negative potential that always dominates over the anisotropies during the contracting phase \cite{Erickson:2003zm} justifying the argument that neglecting the anisotropies can be done in an ekpyrotic scalar field scenario. It has also been shown how one can combine ekpyrotic contraction era and non-singular bounce using a negative exponential potential and a scalar field with a Horndeski-type non-standard kinetic term \cite{Cai_2012}. This model, also discussed in \cite{Cai_2013}, shows how during the entire cosmological evolution of the matter-ekpyrotic bounce, the anisotropies remain small, successfully avoiding the BKL instability. The model is further explored in \cite{Cai:2014zga} via loop quantum cosmology perspective and also in \cite{Cai:2013kja} where the authors have worked with two matter fields, one a scalar field that causes bounce and the other a matter field dominating at the beginning of the contracting phase.

As the model discussed in \cite{Cai_2012,Cai_2013} is free from the above-mentioned instability and is a well-established matter-ekpyrotic bounce scenario, we consider it as our background model. The existence of duality in the scale invariance aspect of the power spectrum \cite{Wands_1999} between an exponentially expanding and collapsing universe motivated us to look for the imprints of anisotropy on the power spectrum if the background is matter-dominated. The imprints caused by the breaking of rotational invariance have been discussed in \cite{Ackerman:2007nb} but considering the matter-dominated collapsing era as a background was not explored yet. We aimed to
search for these imprints in this work. After starting with a period of matter-dominated contraction with
vacuum fluctuations in the massless scalar field, we were able to solve the field equation numerically and find
the quantitative results for the scale invariance part of the power spectrum and the deviation caused by the
presence of anisotropy. 


The outline of this paper is as follows: in the next section, we discuss the model proposed in \cite{Cai_2012,Cai_2013} for a non-singular matter bounce and establish our background as a matter-dominated collapsing universe. Moving to the section \ref{sec:PS}, we work out the power spectrum and the deviations from the scale-invariant part by means of two techniques, one a perturbative approach while the other a direct approach which is discussed in section\ref{sec:PS}\ref{subsec:perturbative_approach} and section\ref{sec:PS}\ref{subsec:direct_approach} in detail. Then in  section\ref{sec:PS}\ref{subsec:Numerical_estimates} we do the numerical estimates for the power spectra on the background universe. Then we conclude our work in section\ref{sec:conclusion} and discuss potential future directions.


\section{The Model}\label{sec:Model}
In this section, we give an overview of the work done in \cite{Cai_2012,Cai_2013}, where the authors have developed an effective model that invokes a smooth bounce through the dynamics of a single scalar field and a flat homogeneous geometry. The metric considered here is a flat, homogeneous but anisotropic, namely the Bianchi-I type, which lacks the rotational invariance and is written as:
\[\dd s^2 = -\dd t^2 + a^2(t)\left[e^{2\theta_1(t)}\dd x^2 + e^{2\theta_2(t)}\dd y^2 + e^{2\theta_3(t)}\dd z^2\right], \]
where $e^{\theta_i(t)}$ represents the anisotropy in scale factor. The FRW universe in an Einstein gravity could yield a successful homogeneous, isotropic, non-singular bounce by violating the Null Energy condition\cite{Cai_2013}.
\medskip
The Lagrangian for such a universe filled with a scalar field $\phi$ and a matter fluid component is \\
\begin{equation}
\mathcal{L}[\phi(x)] = M_p^2\left(1-g(\phi)\right)X + \beta X^2 - V(\phi) + \gamma X \Box\phi ,
\end{equation}
here, $M_p^2$ is the Planck mass square defined to be $1/8\pi G$.\\
The parameter $\beta$ is  positive and bounds the kinetic term from below during high-energy processes, thereby preventing a ghost condition during contraction. The Galileon type operator $\gamma X$ takes care of the gradient instabilities that could creep in due to ghost condensation. The second order derivative in time would be able to establish the ghost condition for a bounce. Nonetheless, the equation of motion obtained from such a Lagrangian is second order. The Null Energy Condition (NEC) is achieved from a negative kinetic part of the equation of motion for $\phi$. The dimensionless function $g(\phi)$ is given by\\
\begin{align}
    g(\phi) &= \frac{2g_0}{e^{-\sqrt{\frac 2 p}\phi} + e^{b_g\sqrt{\frac 2 p}\phi}} ,
\end{align}
 and serves to violate the NEC when approaching bounce $(\phi\rightarrow 0)$, for the non-singular case. Here, $g(\phi)$ has to dominate in the quadratic kinetic term 
 to satisfy for a phase of ghost condensation, which leads to this violation of NEC just before the bounce. The function $g$ starts with very small values for large $\phi$ and attains unity approaching the bounce ($\phi\rightarrow0$). The particular potential
\begin{align}
      V(\phi) &= -\frac{2V_0}{e^{-\sqrt{\frac 2 q}\phi} + e^{b_V\sqrt{\frac 2 q}\phi}},
\end{align}
 is for an Ekpyrotic contraction to follow. The exponential potential $V(\phi)$ is always negative for positive $V_0$. It allows for an attractor behaviour in a contracting cosmology by allowing trajectories of the space-time evolve into one region. The initial condition, $\phi_0$ could be taken from an arbitrary point far in the past, to be an asymptotically large negative value. The potential thus forces the scalar field towards the bounce with $\dot\phi>0$. Also the equation of state is tuned as such to allow an attractor solution namely, $\dot\rho_{\phi}> \dot\rho_m, \dot\rho_{\theta}$. 

\subsection{Background evolution}
The canonical kinetic term in the Lagrangian is \[X = \frac12\partial_{\mu}\phi\partial^{\mu}\phi = \frac12 \dot\phi^2.\] 
And a second kinetic term,
\begin{align*}
 \Box\phi&\equiv \frac{1}{\sqrt{-g}}\partial_{\mu}(\sqrt{-g}g^{\mu\nu}\partial_{\nu}\phi)\\
 &=\ddot\phi + 3H\dot\phi   ,
\end{align*}
for a homogeneous background. For a minimally coupled action,\\
\begin{equation}
    S = \int \mathrm{d}^4x \sqrt{-g}\left[\underbrace{R}_{S_g} + \underbrace{K(\phi,X) + G(\phi,X)\Box\phi}_{S_m} \right],
\end{equation}
varying with respect to the metric gives the energy momentum tensor,
\begin{align}
T_{\mu\nu} \equiv \frac{2}{\sqrt{-g}}\frac{\delta S}{\delta g^{\mu\nu}},
\end{align}
\begin{dmath}\label{eq:2.8}
\Rightarrow T_{\mu \nu}^{\phi}= g_{\mu\nu}\left[V(\phi) - \frac12(1-g(\phi))M_p^2\dot\phi^2 - \frac14\beta\dot\phi^4 + \gamma\dot\phi^2\ddot\phi \right] +\nabla_{\mu}\phi\nabla_{\nu}\phi\left[(1-g(\phi))M_p^2 + \beta\dot\phi^2 + \gamma(\ddot\phi+\sum_i(\dot\theta_i+H)\dot\phi)\right] - \gamma\left[(\nabla_{\mu}\nabla_{\lambda}\phi)\nabla^{\lambda}\phi\nabla_{\nu}\phi + \nabla_{\mu}\phi(\nabla_{\nu}\nabla_{\lambda}\phi)\nabla^{\lambda}\phi\right].
\end{dmath}
For homogeneous universe, the energy momentum tensor can be written in the form of an ideal fluid,
\begin{equation}
    T_{\mu\nu}=(\rho+p)u_{\mu}u_{\nu} + pg_{\mu\nu},
\end{equation}
$u_{\mu}$ being the 4-velocity of the isotropic fluid and $\rho,p$ the respective density and pressure of the fluid.
The corresponding scalar field energy density and pressure for the model under consideration becomes
\begin{align}
\rho_{\phi}&=\frac 1 2 M_p^2(1-g)\dot\phi^2  + \frac 3 4 \beta\dot\phi^4 + 3\gamma H \dot\phi^3 + V(\phi),\label{eq:eqrho}\\
p_{\phi}&= \frac 1 2 M_p^2(1-g)\dot\phi^2 + \frac1 4 \beta\dot\phi^4 - \gamma\dot\phi^2\ddot\phi -V(\phi). \label{eq:eqp}
\end{align}
From the Einstein equations could be worked the equation of motion for the scale factor. The temporal field equation gives the 1st Friedmann equation
\begin{equation}\label{eq:F1}
    H^2 = \frac{\rho}{3M_p^2}+ \frac16\sum_i\dot\theta_i^2,
\end{equation}
while the simplification of the spatial component yields the acceleration equation:
\begin{align}\label{eq:F2}
\dot H = -\frac{\rho_{\phi} + p_{\phi}}{2M_p^2} - \frac12\sum\dot\theta_i^2  \quad \text{and}& \quad\ddot\theta_i + 3H\dot\theta_i= 0.
\end{align}
Inferring from the constraint equation for $H$ \eqref{eq:F1} the anisotropic stress energy density could be expressed as $\sum \dot\theta^2$. The dynamics of whom is according to equation of motion for $\theta_i$ \eqref{eq:F2}. Now solving for $\dot\theta_i$,
\begin{align}
    \dv{\dot\theta_i}{t} &= -3\frac{\dd a}{a\dd t}\dot\theta_i \nonumber\\
    \Rightarrow \dot\theta_i &=M_{\theta,i}\frac{a_B(t)^3}{a(t)^3}, \label{eq:thetadot}
\end{align}
gives the form of the anisotropic energy density, $\dot\theta^2 \propto a^{-6}$ responsible for the BKL instability. Here the parameter $M_{\theta,i}$ is introduced to tune the bounce phase.\\

\subsection{Matter Contraction}
In this model, the universe undergoes a matter-dominated contraction from negative infinity until the slow Ekpyrotic contraction, beginning at $-t_E$ and extending into the bounce.
The mean scale factor here evolves as a power law contraction, given by ,
\begin{equation}\label{eq:meana}
    \Bar{a}(t)\simeq a_E\left(\frac{t-\tilde{t}_E}{t_E-\tilde{t}_E}\right)^{2/3},
\end{equation}
wherein $a_E$ is the transition mean scale factor. The conformal time in the matter bouncing era i.e. from $-t$ to $-t_E$ is:
\begin{equation}\label{conf_time_def}
    \tau=\int_{-t}^{-t_E}\frac{\dd t}{\Bar a(t)}=3(t_E^{1/3}-t^{1/3}).
\end{equation}
and the  mean Hubble parameter is
\begin{equation}
    \Bar{H}(t)=\frac{2}{3(t-\tilde t_E)}.
\end{equation}
In the above $\tilde t_E$ is introduced as an integration constant to match the mean Hubble parameter at transition, $H_E$.
\begin{equation}
    \tilde t_E \simeq t_E -\frac{2}{3H_E}.
\end{equation}
Hence \eqref{eq:meana} allows \eqref{eq:thetadot} to be rewritten as,
\begin{equation}\label{eq:thetadot2}
    \dot\theta_i(t)=M_{\theta,i}\frac{a_B^3\Bar H^2(t)}{a_E^3H_E^2}
\end{equation}
It can be integrated to get the anisotropy factors,
\begin{equation}\label{eq:theta}
    \theta_i(t)=-M_{\theta,i}\frac{2a_B^3\Bar H(t)}{3a_E^3 H_E^2}.
\end{equation}
Here, the mean scale factor during bounce is normalized to unity.

Now, within the set up of this matter contracting bouncing universe, we first only consider a massless scalar field $\chi({\bf x},t)$ which includes the small fluctuations around the background field $\chi_I({\bf x},t)$ and then obtain the power spectrum by computing $\langle\chi(\textbf{x},t)\chi(\textbf{y},t)\rangle$ correlations. We do so via two techniques described in the next section in details. 

\section{Power Spectrum in Matter Dominated Contracting Phase}\label{sec:PS}

To calculate the power spectrum in the matter-dominated contracting universe and search for the existence of the duality in the first-order correction, the first approach that we follow  is the perturbative one. In this, we first approximate the anisotropic metric with a fictitious isotropic metric by considering the deviation from isotropy to be very small. That isotropic metric is the background metric representing the matter-dominated contracting universe. Since the deviation from the isotropy is very small, we can write the scale factors of the anisotropic metric as a small perturbation to the scale factor of the fictitious isotropic metric. Then we define the anisotropy parameter and use it to separate out the scale-invariant part of the power spectrum and the deviation caused by the anisotropy in that. The obtained results are then shown in Table\ref{table:Table_1}, Fig.\ref{fig:delta Pk} and Fig.\ref{fig:ratio delta k log scale}.

The second approach discussed below is a direct one in which we take the free massless field $\chi({{\bf x},t})$ in the anisotropic background, which is still represented as a small perturbation to the fictitious metric. Then we proceed to solve the field equation numerically in the Fourier space and obtain the power spectrum by working out the correlation function $\langle\chi(\textbf{x},t)\chi(\textbf{y},t)\rangle$. Though via this method, we get a complete power spectrum that includes the isotropic part as well as the effect of anisotropy on that. Then we separate out the imprints of the anisotropy and represent those in Fig.\ref{fig:ratio delta Pk}.  

\subsection{Perturbative Approach} \label{subsec:perturbative_approach}
As is discussed in \cite{Ackerman:2007nb}, introducing the anisotropy in the background manifests as direction dependency in the power spectrum. Using the primordial density perturbations $\delta({\bf k})$, the power spectrum is defined as 
\begin{equation}
    \langle\delta(\textbf{k})\delta*(\textbf{q})\rangle=P(k)\delta^3({\bf k - q}).
\end{equation}
Here P(k) may change the form to $P'(k)$ assuming the presence of broken rotational invariance during the inflationary era and can be written in parametric form as,
\begin{equation}\label{eq:ani_ps}
    P'(k)=P(k)\left(1+g(k)(\hat{\textbf{k}}\cdot\textbf{n})^2\right),
\end{equation}
with the line element for the universe characterised by a bidirectional isotropy given by
\begin{equation}\label{eqn:aniso_metric}
    \mathrm{d}s^2 = \mathrm{d}t^2 - a'(t)^2[\mathrm{d}x^2 + \mathrm{d}y^2] - b^2(t)\mathrm{d}z^2.
\end{equation}
Now with the model under consideration as the background, the respective scale factors evolve as
\begin{equation}
   a'(t) = \Bar{a}(t)e^{\theta_1(t)}, 
   \qquad b(t) =\Bar{a}(t)e^{\theta_2(t)}
\end{equation}
and the Hubble parameters become
\[H_a = \frac{\dot a'}{a'},\quad H_b=\frac{\dot b}{b}.\]
The line element written in (\ref{eqn:aniso_metric}) can be approximated using a fictitious isotropic metric and is given by 
\[\mathrm{d}s^2 = \mathrm{d}t^2 - \Bar{a}(t)^2[\mathrm{d}x^2 + \mathrm{d}y^2 + \mathrm{d}z^2],\]
wherein the average scale factor $(\Bar{a}(t))$ evolves as in matter dominated contracting universe case. 
The average Hubble parameter is  $\Bar{H}=\dot{\Bar{a}}/{a}$
and the deviation from isotropy is parametrized using $\epsilon_H$ as \[\epsilon_H=\frac{2(H_b-H_a)}{3\Bar{H}} \simeq \frac{-2}{3}\frac{\Bar{H}}{a_E^3 H_E^2}.\]
Approximating the correlation function in series of $\epsilon_H$, by treating it as a small perturbation \cite{Weinberg_2005}, we can write 
\begin{dmath}\label{eq:corr_func}
    \langle\chi(\textbf{x},t)\chi(\textbf{y},t)\rangle\equiv \langle\chi_I(\textbf{x},t)\chi_I(\textbf{y},t)\rangle + \iota\int_0^t dt'\langle[H_I(t'),\chi_I(\textbf{x},t)\chi_I(\textbf{y},t)]\rangle
\end{dmath}
and the  interaction picture field as a Fourier series in ladder operators as \cite{Weinberg_2005},
\begin{equation}\label{eq:int_field}
    \chi_I(x,t)=\int\frac{\dd^3k}{(2\pi)^3}\left(e^{\iota\textbf{k}\cdot\textbf{x}}\chi(t)a_k + e^{-\iota\textbf{k}\cdot\textbf{x}}\chi^{\ast}(t) a^{\dagger}_k\right).
\end{equation}
The Fourier transform of the two-point function \eqref{eq:corr_func} results in the power spectrum (of the form \eqref{eq:ani_ps}),
\begin{equation}\label{eq:power spectrum}
    \langle\chi(\textbf{x},t)\chi(\textbf{y},t)\rangle=\int \frac{\dd^3k}{(2\pi)^3}e^{\iota\textbf{k}\cdot(\textbf{x}-\textbf{y})}\left(P(k) + (\hat{\textbf k}\cdot\textbf{n})^2\Delta P(k)\right).
\end{equation}
The interaction picture Hamiltonian for a massless scalar field would be,
\begin{equation}\label{eq:int_Hamil}
    H_I(t)=\int \dd^3x\frac12\left[(b-\Bar{a})\left(\dv{\chi}{x_{\perp}}\right)^2 + \left(\frac{a^{\prime2}}{b}-\Bar{a}\right)\left(\dv{\chi}{x^3}\right)^2\right],
    \end{equation}
using the Lagrangian density,
\[\mathcal{L}_{\chi} =-\frac12\sqrt{-g}g^{\mu\nu}\partial_{\mu}\chi\partial_{\nu}\chi,\]
in which we employ the procedure of finding the conjugate momentas for the foreground and background metrics, $\Pi(\chi) = \Pi(\chi)=\pdv{\mathcal{L}(\chi,\partial_{\mu}\chi)}{(\partial_{\mu}\chi)}$ and then the respective Hamiltonian densities $\mathcal{H}[\chi,\Pi] = \dot\chi\Pi - \mathcal{L}_{\chi}$. The interaction part Hamiltonian density, $\mathcal{H}_I$ is the difference in the background from the perturbative densities. \\

Using \eqref{eq:int_field} and \eqref{eq:int_Hamil} in \eqref{eq:corr_func}, and furthermore transforming into the conformal time of the background metric we obtain $P(k)\simeq|\chi_k(\tau)|^2$, and
\begin{align}\label{eq:PS}
        \Delta P(k) &\simeq -6\iota k^2\int_{3t_E^{1/3}}^{\tau}\dd\tau' \Bar{H}\Bar{a}^2(\tau')\epsilon_H(\tau')\bigg(t_E^{1/3}-\frac{\tau'}{3}\bigg)^{5}\left[(\chi_k^{(0)}(\tau')\chi^{(0)\ast}_k(\tau))^2 - (\chi^{(0)\ast}_k(\tau')\chi_k^{(0)}(\tau))^2\right],
\end{align}
where \[\chi_k^{(0)}(\tau)=\frac{\Bar H}{\sqrt{2k}}e^{\iota k \tau}\left[\tau -\frac\iota k\right],\]
becoming
,
\begin{equation}\label{eq:PSint}
    \begin{split}
        \Delta P(k) &\simeq \frac{3a_E}{2k}\left(\frac 2 3 H_E\right)^{-2/3}\int_{3t_E^{1/3}}^{\tau}\dd\tau' \Bar{H}^5\epsilon_H\left[{\left(t_E^{1/3}-\frac{\tau'}{3}\right)}^3 -t_E +\frac 2 3 H_E\right]^{2/3}\\
        &\qquad\times\bigg(t_E^{1/3}-\frac{\tau'}{3}\bigg)^{5}\bigg[(k^2{\tau'}^2-1)\sin{(2k\tau')}+2k\tau'\cos{(2k\tau')}\bigg],
    \end{split}  
\end{equation}
To investigate the scale independent part of the power spectrum and the deviation from it, we multiply by $k^3$ and get: 
\begin{equation} \label{pk_def_I_approach}
P'(k)k^3 = P(k)k^3 + (\hat{\textbf k}\cdot\textbf{n})^2 \Delta P(k)k^3,
\end{equation}
where,
\begin{align}
    P(k)k^3 &= \frac{\Bar{H}^2(\tau)}{2} , \label{eqn:iso_pow_spec}\\
     \Delta P(k)k^3 &\simeq \frac{3a_Ek^2}{2}\left(\frac 2 3 H_E\right)^{-2/3}\int_{3t_E^{1/3}}{}^{\tau}\dd\tau' \Bar{H}^5\epsilon_H{\left[{\left(t_E^{1/3}-\frac{\tau'}{3}\right)}^3 -t_E +\frac 2 3 H_E\right]}^{2/3}\\\nonumber
        &\qquad\times\bigg(t_E^{1/3}-\frac{\tau'}{3}\bigg)^{5}\bigg[(k^2{\tau'}^2-1)\sin{(2k\tau')}+2k\tau'\cos{(2k\tau')}\bigg], 
\end{align}

\subsection{Direct Approach with Scalar Field in Anisotropic Background} \label{subsec:direct_approach}
Another way to obtain the power spectrum is to solve the field equation $\Box\chi= 0$ for a massless scalar field directly in the Fourier space and then working out the correlation function using the solutions obtained. This equation
for the modes with wavenumbers along the $\hat{z}$ direction in the anisotropic background can be written in the Fourier space as 
\begin{equation}
    \dv[2]{\chi_k}{t} + 3\Bar{H}\dv{\chi_k}{t} + \frac{k^2}{\Bar{b}(t)^2}\chi_k=0 ,
\end{equation}
which in conformal time becomes
\begin{equation}\label{conf_field_eqn}
    \dv[2]{\chi_k}{\tau}\frac{1}{Y(\tau)^4} + \left[\frac{2}{3Y(\tau)^5}-\frac{3\Bar{H}(\tau)}{Y(\tau)^2}\right]\dv{\chi_k}{\tau} + \frac{k^2}{\Bar{b}(\tau)^2}\chi_k=0 ,
\end{equation}
where,
\begin{align*}
Y(\tau)&=\left(t_E^{1/3}-\frac{\tau}{3}\right),\\
H(\tau)&=\frac{2}{3\left[Y(\tau)^3-t_E+2/3 H_E\right]},\\
b(\tau)&=a_E\left[\frac{Y(\tau)^3-t_E+2/3 H_E}{2/3 H_E}\right]\\&\qquad\times\exp{-M_{\theta,2}\frac{2}{3a_E^3H_E^2}\Bar{H}(\tau)},
\end{align*}
For simplicity, in dealing with scale factors, the parametrization
\[b(\tau) = \Bar{a}(\tau)\left(1 + \epsilon_H\right),\]
 has been used. For solving eqn.(\ref{conf_field_eqn}), we set up all the parameters in the next section for the matter dominated background case and find the power spectrum numerically using both the approaches.
 
 \subsection{Numerical Estimates of Power Spectrum with Both Approaches}\label{subsec:Numerical_estimates}
 
To numerically evaluate the power spectrum and its first order correction, we work in the same parameter regime as \cite{Cai_2013} and write all relevant parameters and functions in the reduced Planck mass units as
\begin{center}
\begin{tabular}{ c c c c }
 $V_0=10^{-7} M_p^4$, & $g_0 =1.1$, & $\beta=5$, & $\gamma=10^{-3}$,\\
 $b_V=5$, & $b_g=0.5$, & $p=0.01$, & $q=0.1$  ,
\end{tabular}
\end{center}

 and then from Hubble parameter vs time plot (Figure 2) and energy density vs time (Figure 3) in \cite{Cai_2013}, we get
\begin{equation}
        t_E=-2.486\times10^{3},\quad H_E=-6.1\times10^{-6},
\end{equation}
 which is required for our model. Also, we set $a_E=1$ and $\epsilon_H=10^{-5}$ and the range of the conformal time ($\eta$) for the aforementioned parameter regime turns out to be from $-9.6$ to $0$.
 
\begin{figure} [H]
    \centering
    \includegraphics[width=150mm]{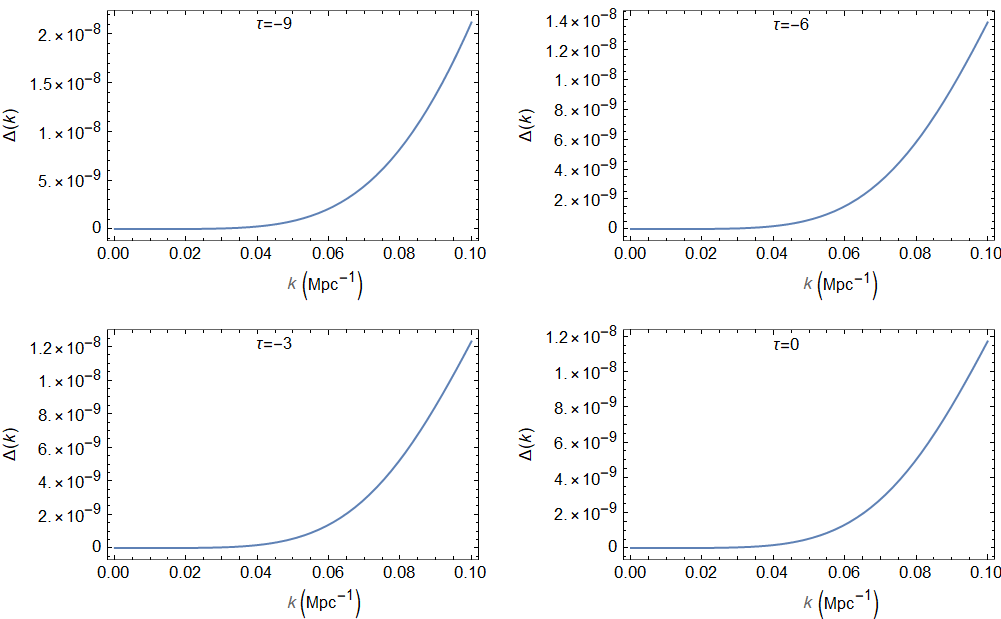}
    \caption{ $\Delta (k)$ vs $k$ using the perturbative method at four different values of $\tau$.}
    \label{fig:delta Pk}
\end{figure}

\begin{figure} 
    \centering
    \includegraphics[width=150mm]{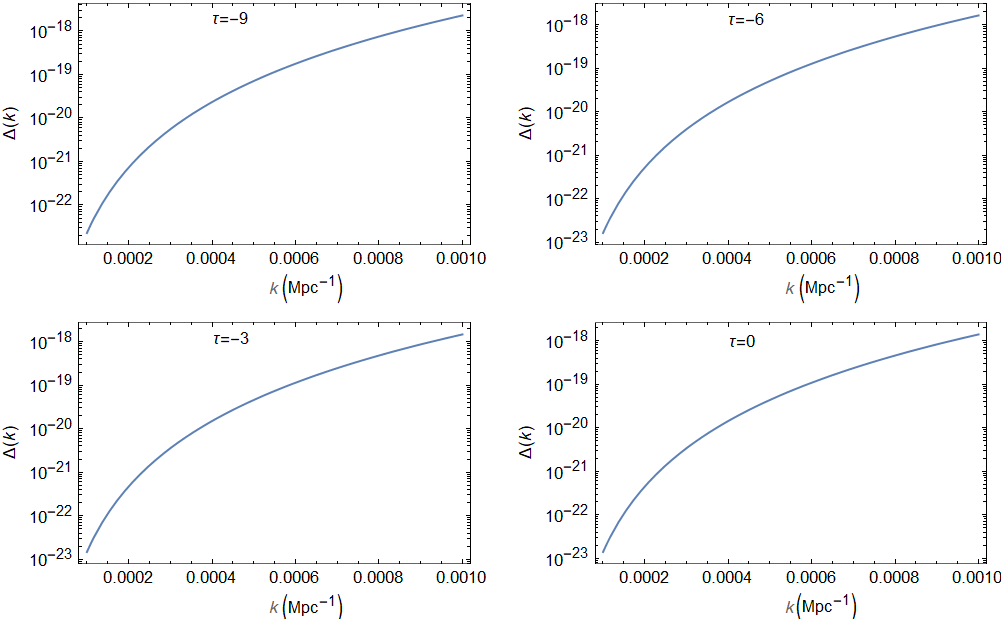}
    \caption{$\Delta (k)$ vs $k$ for four different values $\tau$ on log scale via perturbative technique.}
    \label{fig:ratio delta k log scale}
\end{figure}

 
 With the perturbative approach, we get the results shown in Table \ref{table:Table_1},  Fig.\ref{fig:delta Pk} and Fig.\ref{fig:ratio delta k log scale}. The Table \ref{table:Table_1} consists the values for isotropic part of the power spectrum corresponding to each value of the conformal time $(\tau)$. As is evident there itself, this is approximately scale invariant with being independent of wavenumber $k$ as defined in eqn.(\ref{eqn:iso_pow_spec}) and is of the order of $10^{-10}$. Next in Fig.\ref{fig:delta Pk},  we have plotted $\Delta(k)$ vs $k$ where $\Delta(k)=\Delta P(k)/P(k)$ at four different values of conformal time for the range of k from $10^{-4} Mpc^{-1}$  to $10^{-1} Mpc^{-1}$ as this is the regime over which the measurement of primordial power spectrum is done by studying the fluctuation and anamolies in CMB \cite{WMAP:2006bqn,Copi:2006tu,Bennett:1996ce,WMAP:2006jqi,Planck:2019evm,Rubtsov:2014yua}. Here, $\Delta(k)$ is found to be approximately within order of $10^{-8}$ and  $10^{-9}$  for the entire range of the conformal time in our model and has these significant values for the modes only lying between $4\times 10^{-2} Mpc^{-1}$ to $ 10^{-1} Mpc^{-1}$. Then in Fig.\ref{fig:ratio delta k log scale}, we presented the results for smaller k modes from $10^{-4} Mpc^{-1}$ to  $10^{-3} Mpc^{-1}$ on a log scale. Here, $\Delta(k)$ is found to be much smaller than the observed values of the anomalies from CMB and lies in the range of $10^{-23}$ to $10^{-18}$.
 
 \begin{table}[h!]
 \begin{center}
\begin{tabular}{||c c c ||} 
 \hline
 S.No. & Conformal time & Isotropic Power Spectrum \\ [0.5ex] 
 \hline\hline
 $1$ & $-9.0$ & $2.41968\times10^{-10}$  \\ 
 \hline
 $2$ & $-6.0$ & $2.56600\times10^{-10}$  \\
 \hline
 $3$ & $-3.0$ & $2.70710\times10^{-10}$  \\
 \hline
 $4$ & $0$ & $2.84106\times10^{-10}$  \\ [1ex]
 \hline
\end{tabular} 
\caption{Conformal time and corresponding values of the isotropic power spectrum.}
 \label{table:Table_1}
\end{center}
 \end{table}
 
 
\begin{figure}[H]
    \centering
    \includegraphics[width=150mm]{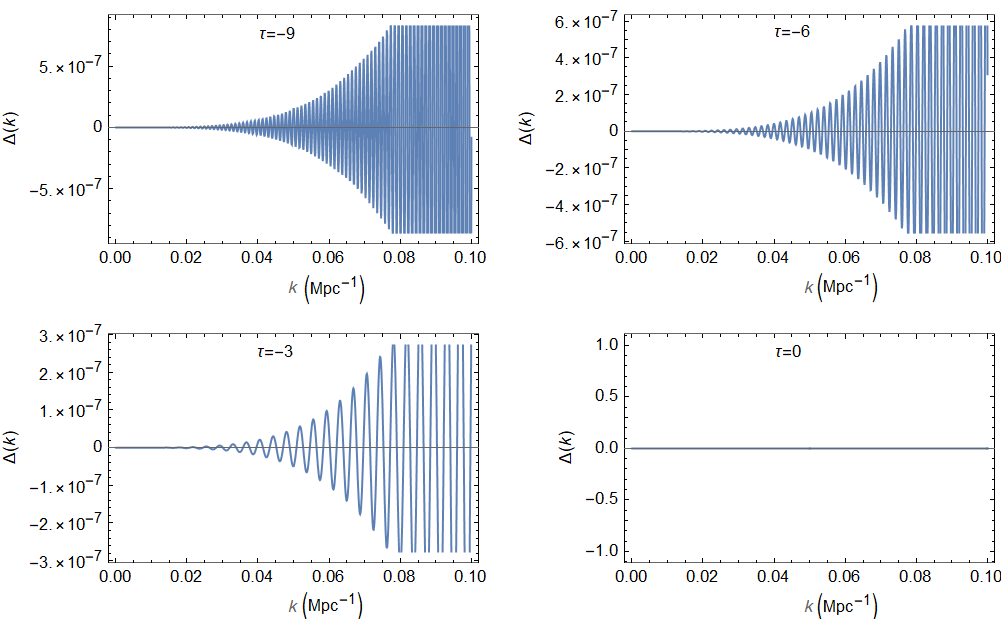}
    \caption{ $\Delta (k)$ vs $k$ using the direct method at four different values of $\tau$.}
    \label{fig:ratio delta Pk}
\end{figure}

Via the second technique, we get the results shown in Fig.\ref{fig:ratio delta Pk} for the power spectrum. For this,  we first numerically solved eqn.(\ref{conf_field_eqn}) for $\chi_k$ and then plotted $\Delta(k)$ vs $k$. The amplitude of $\Delta(k)$ is of the order $10^{-7}$, and not much variation can be seen with k ranging from $4 \times 10^{-2} Mpc^{-1}$ to $10^{-1} Mpc^{-1}$ as was the case in the first approach. Though the results from both the approaches do not match exactly, that might be because of the approximation not working for this parameter regime but this helps us in setting up an upper bound on $\Delta(k)$ which is of $10^{-7}$ order. The other difference in the results using the second approach is an oscillatory character, as seen in Fig.\ref{fig:ratio delta Pk}. It is because, in the first approach, the oscillations are absorbed in the ${\bf k\cdot n}$ factor, as shown in eqn.(\ref{pk_def_I_approach}). Since $\tau=0$ is the final limit of the conformal time we have considered, we get the trivial result of the field equation within the aforementioned range of wavenumber k, zero. Hence, our results from both techniques are expected not to match at this value of conformal time, and the same is verified by the numerical results.


\section{Discussion and Conclusion}\label{sec:conclusion}

\par
In this paper, we have worked out quantitatively the imprints that the anisotropy can cause on the scale-invariant power spectrum, given the universe had a matter-dominated contracting phase as the background. The argument for considering this as the background comes from \cite{Wands_1999}, in which the author showed that there exists a duality between the two methods of calculating the power spectrum of perturbations for a minimally coupled massless field. The first is the exponentially expanding universe in which the scale factor goes as $a\propto e^{Ht}$, while the second is a matter-dominated collapsing universe with $a(t)\propto (-t)^{2/3}$. The first case has been explored in \cite{Ackerman:2007nb} with detailed discussions on imprints of the anisotropy and breaking of rotational invariance but, the dual to the former one had not been explored yet, thus leading to this work. Using the method described in \cite{Cai_2013}, first we establish the matter-dominated contracting universe as a background for exploration. Then, within the same regime of parameters used in \cite{Cai_2013}, we worked out the power spectrum by studying the correlations for a massless scalar field in this background and looked at the effects that the anisotropy can cause if it is present from the beginning. To study those imprints, we have used two techniques first, the perturbative approach, while the second technique was the direct approach. The quantitative results for the deviation from the scale-invariant power spectrum were obtained using both methods for the range of k between $10^{-4} Mpc^{-1}$ to $10^{-1} Mpc^{-1}$, as these are the modes of interest based on cosmological observations as discussed earlier. The $\Delta(k)$ is found to be of order of $10^{-8}$ and $10^{-9}$ via the perturbative approach, while the directly solving the field equation and evaluating the correlation gave us $\Delta(k)$ of the order of $10^{-7}$. The reason for both the values to be different might have to do with the parameters regime under consideration over here.  Though the deviation was found to be inconsistent with the recent observation from Planck data in \cite{Planck:2019evm}, the work done here helps us get an upper bound (approximately of the order of $10^{-7}$) on the imprints for the aforementioned range of k modes. 
Also, we observed in our results that the $\Delta(k)$ has significant values for the modes lying between $4 \times 10^{-2} Mpc^{-1}$ to $10^{-1} Mpc^{-1}$ via both the approaches. These values can further be constrained if one uses other sets of parameters maintaining the stability of the matter bounce scenario. On exploring more for much smaller modes in the range $10^{-4} Mpc^{-1}$ to $10^{-3} Mpc^{-1}$, we found $\Delta(k)$ varying between $10^{-23}$ to $10^{-18}$, which is much smaller than the observed values of anomalies in CMB. Since the value for $\Delta(k)$ differs from the one we get via the inflationary scenario, the work done in this paper can help us in distinguishing between both matter bounce and inflation via looking at just the power spectra, but mismatching of the results between both implies that matter bounce may not be a good candidate for explaining CMB anomalies at low k, however, it would be interesting to do a similar study for other bouncing models. This work is further left for exploration via CMB constraints and calculation of other cosmological parameters. We plan to do those in future.

\section{Acknowledgement}
This work is partially supported by DST
(Govt. of India) Grant No. SERB/PHY/2021057.
\begin{center}
 \rule{4in}{0.5pt}\\
 \vspace{-11.5pt}\rule{3in}{0.5pt}\\
 \vspace{-11.5pt}\rule{2in}{0.5pt}\\
\end{center}
\printbibliography[heading=none]
\end{document}